\documentclass[showkeys,aps,twocolumn,showpacs,preprintnumbers,amsmath,amssymb,prl,letterpaper,floatfix,nofootinbib,superscriptaddress,]{revtex4-1}

\usepackage{amsmath,amssymb}
\usepackage[usenames,dvipsnames]{color}
\usepackage{graphicx}
\usepackage{subfigure}
\usepackage{soul}
\newcommand\bef{\begin{figure}}
\newcommand\eef[1]{\label{fg:#1}\end{figure}}
\newcommand\beq{\begin{equation}}
\newcommand\eeq[1]{\label{#1}\end{equation}}
\newcommand\bet{\begin{table}}
\newcommand\eet[1]{\label{tb:#1}\end{table}}

\usepackage{hyperref}
\hypersetup{colorlinks, linkcolor = [rgb]{0,0.0,0.75}, citecolor = [rgb]{0,0.0,0.75}, urlcolor = [rgb]{0,0.0,0.75}}

\begin{document}
\title{Deuteron-like heavy dibaryons from Lattice Quantum Chromodynamics}

\author{Parikshit\ \surname{Junnarkar}}
\email{parikshit@theory.tifr.res.in}
\affiliation{Department of Theoretical Physics, Tata Institute of Fundamental
         Research,\\ Homi Bhabha Road, Mumbai 400005, India.}

\author{Nilmani\ \surname{Mathur}}
\email{nilmani@theory.tifr.res.in}
\affiliation{Department of Theoretical Physics, Tata Institute of Fundamental
         Research,\\ Homi Bhabha Road, Mumbai 400005, India.}

\pacs{12.38.Gc, 12.38.-t, 14.20.Lq}

\begin{abstract}
We report the first lattice quantum chromodynamics (QCD) study of
deuteron($np$)-like dibaryons with heavy quark flavours. These include
particles with following dibaryon structures and valence quark
contents: $\Sigma_c\Xi_{cc} (uucucc)$, $\Omega_c\Omega_{cc} (sscscc)$,
$\Sigma_b\Xi_{bb} (uububb)$, $\Omega_b\Omega_{bb} (ssbsbb)$ and
$\Omega_{ccb}\Omega_{cbb} (ccbcbb)$, and with spin ($J$)-parity ($P$),
$J^{P} \equiv 1^{+}$. Using a state-of-the art lattice QCD
calculation, after controlling relevant systematic errors, we
unambiguously find that the ground state masses of dibaryons
$\Omega_c\Omega_{cc} (sscscc)$, $\Omega_b\Omega_{bb} (ssbsbb)$ and
$\Omega_{ccb}\Omega_{cbb} (ccbcbb)$ are below their respective
two-baryon thresholds, suggesting the presence of bound states which
are stable under strong and electromagnetic interactions. We also
predict their masses precisely. For dibaryons
$\Sigma_c\Xi_{cc} (uucucc)$, and $\Sigma_b\Xi_{bb} (uububb)$, we could not
reach to a definitive conclusion about the presence of any bound
state due to large systematics associated with these states. We also
find that the binding of these dibaryons becomes stronger as they
become heavier in mass.  This study also opens up the possibility of
the existence of many other exotic nuclei, which can be formed through
the fusion of heavy baryons, similar to the formation of nuclei of
elements in the Periodic Table.
\end{abstract}
\maketitle


A deuteron is a bound state of two baryons, a proton and a neutron, and is made of six light valence quarks. In the early Universe, deuterons were created and their stability is responsible for the creation of other elements. Interestingly, the strong interactions between quarks, which bring stability to deuterons, also allow various other six-quark combinations leading to the possible formation of many other dibaryons. However, no such strong-interaction-stable dibaryons, though speculated about many times \cite{Jaffe:1976ig,Mulders:1980vx,Balachandran:1983dj,Sakai:1999qm,Ikeda:2007nz, Bashkanov:2013cla, Shanahan:2011su, Clement:2016vnl}, have been observed yet \cite{Echenard:2018rfe}. Using a state-of-the-art first principles calculation of lattice quantum chromodynamics (QCD), here we report, for the first time, a definite prediction of the existence of other deuteron-like spin-1 dibaryons. We also predict their masses precisely. These new subatomic particles could either be made of six heavy quarks (charm and bottom) or heavy and strange quarks. Such dibaryons are stable against strong and electromagnetic decays, but, unlike the deuteron, they can decay through weak interactions. We also find that such dibaryons become more strong-interaction-stable as they become heavier.
We expect that prediction from this
calculation will initiate more theoretical works on heavy dibaryons, particularly to understand their binding mechanism, and may as well aid in discovering these new subatomic particles at future experimental facilities, such as at the upgraded Large Hadron Collider and future high energy heavy ion facilities.
In fact this study opens up the possibility of the existence of many other exotic nuclei, which can be formed through the fusion of heavy baryons, similar to the formation of nuclei of elements in the Periodic Table. Formation of these hadrons perhaps also enhances the possibility of a quark-level analogue of nuclear fusion as discussed recently \cite{Karliner:2017elp}. Further study of these exotic states can also provide information on the strong interaction dynamics at multiple scales.

The particular dibaryons ($\mathcal{D}$)\footnote{We identify the two-flavour spin-1 dibaryons with the symbol $\mathcal{D}$ and name them as $\mathcal{D}_{q_1q_2}$, which are made of two baryons with valence quark contents $(q_1 q_1 q_2)$ and $(q_1 q_2 q_2)$. In this notation the deuteron is $\mathcal{D}_{ud}\equiv np(uududd)$.  Such dibaryons may be called as $\mathcal{D}$-dibaryons.}
that we investigate are heavy quark analogues of deuteron ($np$). They have the spin-($J$)-parity ($P$) quantum numbers: $J^P = 1^+$, with following dibaryon configurations: $\mathcal{D}_{cu} \equiv \Sigma_c\Xi_{cc} (uucucc)$, $\mathcal{D}_{cs} \equiv \Omega_c\Omega_{cc} (sscscc)$, $\mathcal{D}_{bu} \equiv \Sigma_b\Xi_{bb} (uububb)$, $\mathcal{D}_{bs} \equiv \Omega_b\Omega_{bb} (ssbsbb)$ and $\mathcal{D}_{bc} \equiv \Omega_{ccb}\Omega_{cbb} (ccbcbb)$. Here, $\Sigma_q,\Xi_{qq},\Omega_{qq},\Omega_{q_1q_2q_2}$'s are heavy baryons with the usual nomenclature of the Particle Data Group \cite{Tanabashi:2018}, and $u,s,c,b$ inside brackets are various quark flavours.  We find that $D_{cs}$, $D_{bs}$ and $D_{bc}$ are stable against strong and electromagnetic decays, and thus it will interesting to carry out further theoretical studies on their binding mechanisms as well as their production mechanisms. However, for $\mathcal{D}_{cu}$ and  $\mathcal{D}_{bu}$,  we find the ground state masses are consistent with their respective two-baryon thresholds, and therefore our results are not currently precise enough to reach a definitive conclusion on their stability.
Incidentally, only recently  tetra-\cite{Choi:2007wga, Aaij:2014jqa, 
  Belle:2011aa} and pentaquark \cite{Aaij:2015tga,Aaij:2019vzc} states have been discovered and those are made of heavy quarks. Recent model and lattice QCD studies also suggest the existence of other heavy tetraquark hadrons
\cite{Karliner:2017qjm, Eichten:2017ffp, Francis:2016hui,Bicudo:2017szl,Junnarkar:2018twb,Francis:2018jyb,Leskovec:2019ioa}. It is thus natural to search for six-quark states made of heavy quarks and our predictions provide first indication for the existence of such heavy hadrons.
Dibaryons have been studied through various models over the years \cite{Jaffe:1976ig,Balachandran:1983dj,Mulders:1980vx, Sakai:1999qm, Ikeda:2007nz, Bashkanov:2013cla, Shanahan:2011su, Clement:2016vnl}, However, it is quite crucial to
have first principles lattice QCD based studies on these states to predict their masses and to understand their structures. In fact a few lattice QCD
studies have already been carried out~\cite{Inoue:2010es,Beane:2010hg,Beane:2011iw,Yamazaki:2012hi,Gongyo:2017fjb,Francis:2018qch,Iritani:2018sra}. However, those are mainly focused on light quarks with spin-0 states \cite{Inoue:2010es,Beane:2010hg,Gongyo:2017fjb,Francis:2018qch}, studies of deuteron in Refs. \cite{Beane:2011iw,Yamazaki:2012hi} as well as studies of spin-2 states \cite{Iritani:2018sra}. This work is the first lattice study on heavy dibaryons. A lattice dibaryon calculation with heavy quarks is advantageous over the light counterparts in two ways.
The two point correlators are less noisy in comparison with light dibaryons and secondly the signal-to-noise ratio of the two point correlators is far better in comparison to the light quark calculations.
Both of these provide an added advantage in our calculation and help in reliable extraction of binding energies of these heavy dibaryons.

The lattice set up that we utilize here is similar to that was used in Refs. \cite{Basak:2012py,Mathur:2018epb, Mathur:2018rwu,Junnarkar:2018twb}. Below we elaborate it further for the sake of completeness.

\noindent{\bf{A. Lattice ensembles:}}
Three sets of dynamical 2+1+1 flavours ($u/d,s,c$) lattice ensembles, within a volume of about 3 fm, generated by MILC collaboration~\cite{Bazavov:2012xda} with HISQ fermion action~\cite{Follana:2006rc} are employed for this study.  Lattice
spacings, using $r_1$ parameter, for these ensembles are measured to be 0.1207(11) 0.0888(8) and 0.0582(5) {\it{fm}}, respectively~\cite{Bazavov:2012xda}, which are also found to be consistent with scales obtained through Wilson flow~\cite{Bazavov:2015yea}.

\noindent{\bf{B. Quark actions:}} In the valence sector, from light to
charm quarks, we utilize the overlap action
which has exact chiral symmetry at finite lattice spacings~\cite{Neuberger:1997fp, Neuberger:1998wv, Luscher:1998pqa} and no $\mathcal{O}^n(ma), n = 1, 3, 5, \cdots$ errors. A non-relativistic QCD (NRQCD)
formulation \cite{Lepage:1992tx} is adapted for the bottom quark with an $\mathcal{O}(\alpha_sv^4)$ improved Hamiltonian with
non-perturbatively tuned improvement coefficients \cite{Dowdall:2011wh}.
For reliable extraction of the ground states we employ a wall source and point sink. 

\noindent {\textbf{C. Quark mass tuning:} We tune both the charm and bottom quark masses using the Fermilab prescription for heavy quarks~\cite{ElKhadra:1996mp} in which we equate the lattice-extracted spin-averaged kinetic masses of the $1S$ quarkonia states with their physical values \cite{Tanabashi:2018}. Following Ref.~\cite{Chakraborty:2014aca} we tune the strange quark mass to its physical value.

  \noindent\textbf{D. Dibaryon interpolators:}
   The dibaryon interpolating operator with spin $(J=1)$ and antisymmetric in flavours $(q,Q) \in (s,c,b)$ is constructed as:
\begin{equation}
{\mathcal{D}}_{qQ} = \frac{1}{\sqrt{2}}  \bigg( \Omega_{qqQ} (C \gamma^j) \Omega_{QQq} - \Omega_{QQq} (C \gamma^j) \Omega_{qqQ}  \bigg)
\end{equation}
   where $\Omega_{qqQ}$ and $\Omega_{QQq}$ are spin-1/2 baryons defined as, $(\Omega_{qqQ})_\alpha = \epsilon^{abc} q^a_\alpha (x) q^b_{\mu}(x) (C\gamma_5)_{\mu \nu} Q^c_{\nu}(x)$ and $(\Omega_{QQq})_\alpha = \epsilon^{abc} Q^a_\alpha (x) q^b_{\mu}(x) (C\gamma_5)_{\mu \nu} Q^c_{\nu}(x)$. Here Latin letters indicate color, while Greek letters indicate the spinor degree of freedom.
   The various deuteron analogues with appropriate flavour antisymmetry are listed in Table~\ref{tb:op}.

In a lattice QCD formulation \cite{Wilson:1974sk}, the mass ($m$) of a particle is extracted in two steps: first by calculating the two-point Euclidean time ($\tau$) correlator ($\langle C(\tau)\rangle$) of the interpolating source and sink operators, over many gauge configurations, and then extracting the exponent ($m$) by fitting the exponential decay ($\langle C(\tau)\rangle \, \sim e^{-m\tau}$) of that correlator at large Euclidean time. Following the dibaryon structure of deuteron ($d = {1\over \sqrt{2}} (pn-np)$), the spin-1 flavour-antisymmetric dibaryon interpolating fields are constructed with the appropriate spin projection of two individual spin-1/2 baryons, as shown in  Table ~\ref{tb:op}.
We then calculate the two-point correlators with those interpolating fields and from the exponential decay of these correlators we calculate the lowest energy states, {\it i.e.}, the ground state masses of each dibaryons. Masses of individual baryons are also calculated similarly (See Supplemental Material [url] for more details on calculation procedure and error analysis, which includes Refs. \cite{Lepage:2001ym,Mathur:2018epb,Mathur:2018rwu,Brown:2014ena,Basak:2013oya,Bazavov:2015yea,ElKhadra:1996mp,Luscher:1990ck,McNeile:2012qf,Dowdall:2012ab,Chakraborty:2014aca,Basak:2014kma,Borsanyi:2014jba}).
\begingroup
\renewcommand*{\arraystretch}{1.9}
\bet[h]
\centering
\begin{tabular}{c|c}\hline \hline 
$\mathcal{D}_{Qq}$ & Interpolating fields \\ \hline \hline
$\mathcal{D}_{bc}$ & $\frac{1}{\sqrt{2}}  \big( \Omega_{ccb} \Omega_{bbc} - \Omega_{bbc} \Omega_{ccb}  \big)$ \\ \hline
$\mathcal{D}_{bs}$ & $\frac{1}{\sqrt{2}}  \big( \Omega_{b} \Omega_{bb} - \Omega_{bb} \Omega_{b}  \big)$ \\ \hline
$\mathcal{D}_{cs}$ & $\frac{1}{\sqrt{2}}  \big( \Omega_{c} \Omega_{cc} - \Omega_{cc} \Omega_{c}  \big)$ \\ \hline
$\mathcal{D}_{bu}$ & $\frac{1}{\sqrt{2}}  \big( \Sigma_{b} \Xi_{bb} - \Xi_{bb} \Sigma_{b}  \big)$ \\ \hline
$\mathcal{D}_{cu}$ & $\frac{1}{\sqrt{2}}  \big( \Sigma_{c} \Xi_{cc} - \Xi_{cc} \Sigma_{c}  \big)$ \\ \hline
\end{tabular}
\caption{\label{tb:op}Structures of spin-1 heavy dibaryons that we study in this work.}
\eet{tb:op}
\endgroup

The next step is to find the relative energy levels of the ground state dibaryons with respect to their two-baryon thresholds.
Since baryon number is a conserved quantity in the Standard Model of particle physics, these spin-1 dibaryons can have only two strong decay channels: either to two spin-1/2 baryons or to two spin-3/2 baryons. Interestingly, we find that the ordering of masses for these two combinations is different with the charm and the bottom quarks. While the sum of masses of two spin-1/2 charmed baryons ($\Sigma_{c}$ and $\Xi_{cc}$ or $\Omega_{c}$ and $\Omega_{cc}$) are smaller than that of two spin-3/2 charmed baryons ($\Omega_{qqq,q=u,s}$ and $\Omega_{ccc}$),  for the bottom quark this trend is opposite, {\it i.e.}, $M_{\Sigma_b} + M_{\Xi_{bb}} > M_{\Delta_{uuu}} + M_{\Omega_{bbb}}$, $M_{\Omega_b} + M_{\Omega_{bb}} > M_{\Omega_{sss}} + M_{\Omega_{bbb}}$ and $M_{\Omega_{ccb}} + M_{\Omega_{cbb}} > M_{\Omega_{ccc}} + M_{\Omega_{bbb}}$. These observations are consistent with known experimental results and lattice determination of baryon masses \cite{Brown:2014ena,Mathur:2018epb, Mathur:2018rwu}. After computing the dibaryon masses we then calculate their mass differences from both the spin-1/2 and spin-3/2 two-baryon thresholds. When the mass difference of a dibaryon from its closest threshold is negative and the finite volume effects are convincingly small, that particular dibaryon is likely to be a bound state.


\bef[t]
\centering
\includegraphics[scale=0.45]{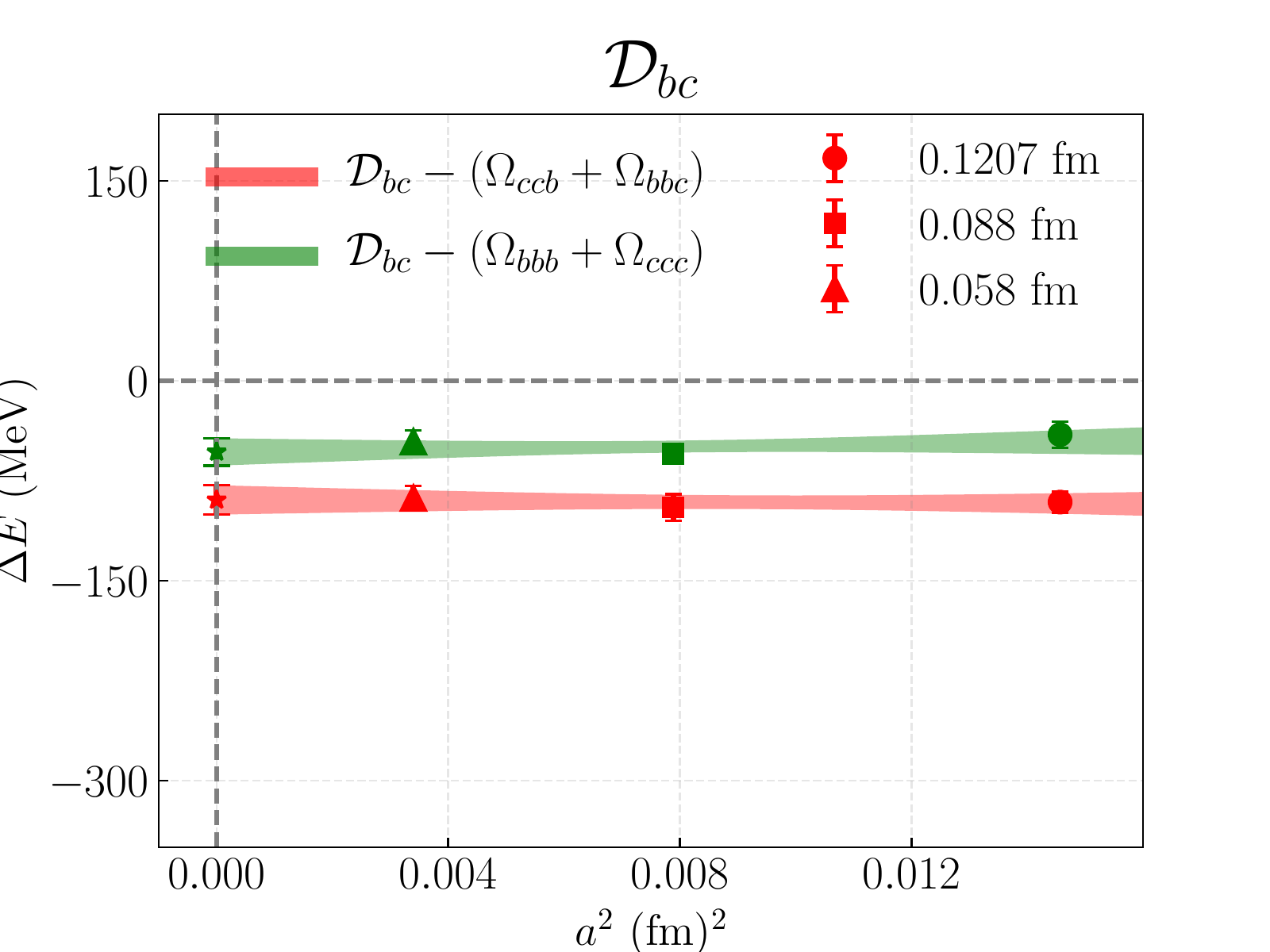}
\includegraphics[scale=0.45]{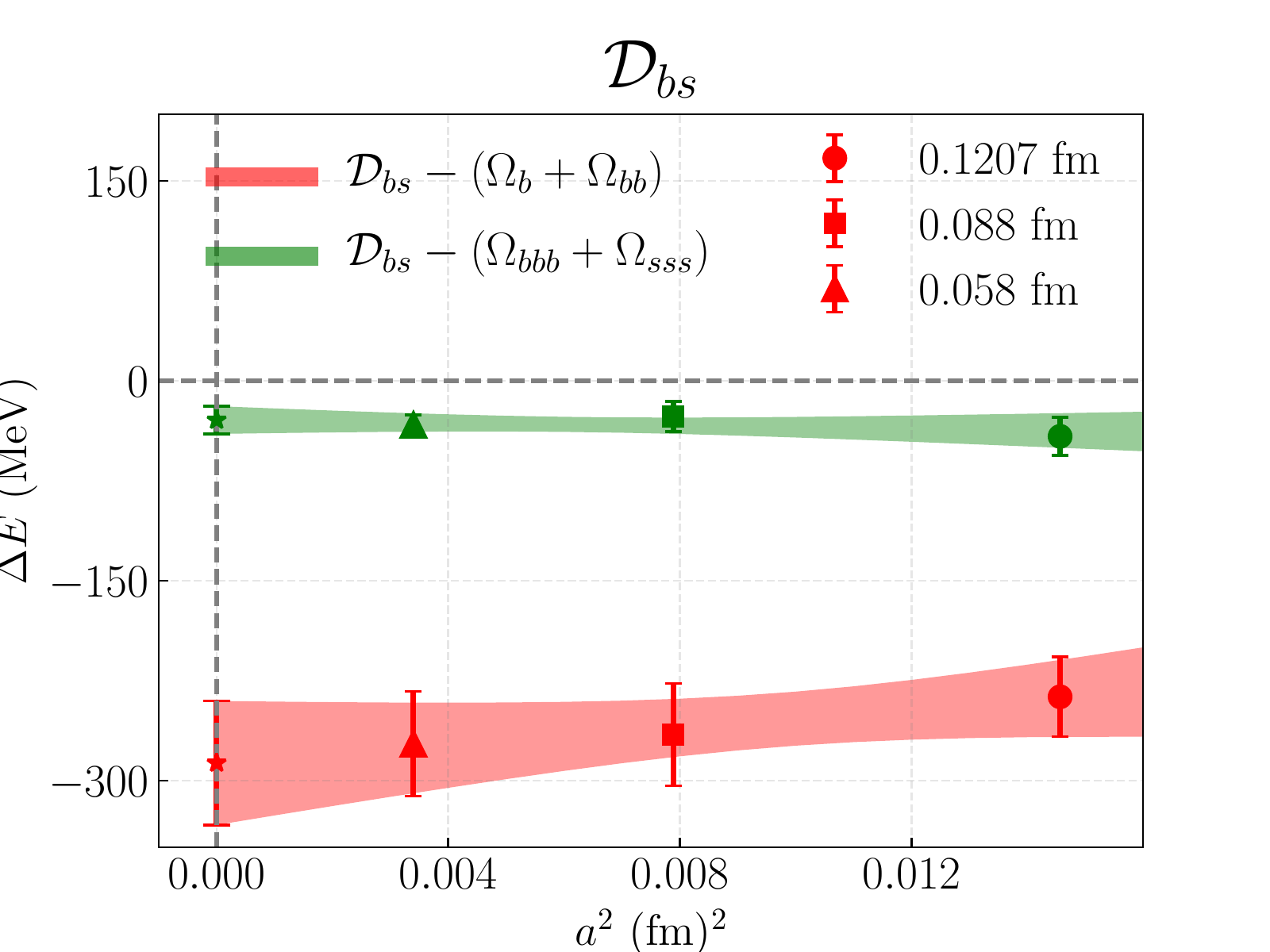}
\includegraphics[scale=0.45]{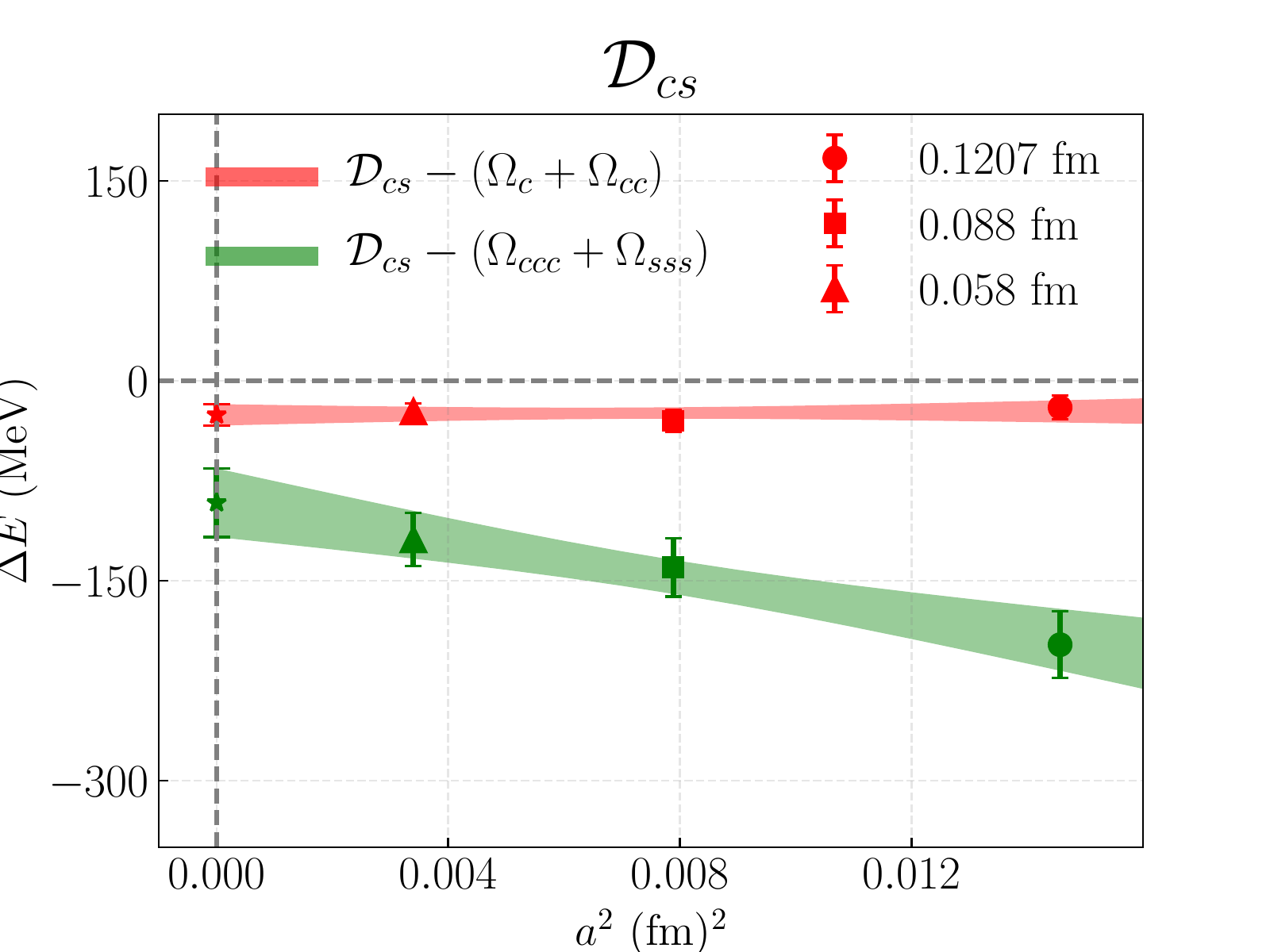}
\vspace*{-0.09in}
\caption{\label{fig:DB_qQ} Mass differences between various spin-1 heavy dibaryons ($\mathcal{D}_{q_1q_2}$) from their two-baryon threshold states. Red points represent mass differences when the threshold states are with spin-1/2 baryons and the green points are the same with spin-3/2 baryons. Results are shown at three lattice spacings and at the continuum limit with shaded bands as one sigma error.}
\eef{}

We compute the aforementioned mass differences at multiple lattice spacings within our lattice set-up. Our results are shown in Figure~\ref{fig:DB_qQ}. Mass differences from spin-1/2 and spin-3/2 thresholds are shown by red and green colours respectively. The upper plot is for the dibaryon $\mathcal{D}_{bc}(\Omega_{ccb}\Omega_{cbb})$, middle one is for $\mathcal{D}_{bs}(\Omega_{ssb}\Omega_{sbb})$ and the bottom one is for $\mathcal{D}_{cs}(\Omega_{ssc}\Omega_{scc})$.
It should be noted that these dibaryon energy levels are already computed at their physical quark masses and therefore only require a continuum extrapolation and finite volume corrections.
For the continuum extrapolation, the energy levels are computed at three lattice spacings, indicated by different marker styles,  and we use a linear in $a^2$  ansatz as well as with an $a^2\,\rm{ln}(a)$  term. To be noted that use of overlap action, which has exact chiral symmetry on lattice, ensures no $\mathcal{O}^n(ma), n = 1, 3, 5, \cdots$ errors, which then assures that higher order discretization errors are smaller, particularly at the finest lattice. In addition, since we calculate mass splittings, rather than masses, errors due to finite lattices are in good control. The errors shown include both statistical as well as systematic errors which are then added in quadrature to get the final errors (see Supplemental Material for details on error analysis).

From these figures it is quite apparent that the ground state masses of $\mathcal{D}_{bc}, \mathcal{D}_{bs}$ and $\mathcal{D}_{cs}$ lie below their respective closest two baryon thresholds by about 4, 2 and 3 sigma errors, respectively. Next,  it is natural to ask if these mass differences obtained at finite volume lattices ($L^3$) are indeed the physical binding energies that hold these dibaryons together from decaying to individual baryons. To sort this out one needs to study finite volume corrections of these lattice computed energy levels. Fortunately for multi-hadron systems this has already been worked out \cite{Beane:2003da,Davoudi:2011md,Briceno:2013bda} and the finite volume corrections, $\Delta_{FV}$, for such systems was found to be $\sim {\mathcal{O}}(e^{-k_{\infty} L})/L$, where $k_{\infty} = \sqrt{(m_1+m_2)B_{\infty}}$, $m_1, m_2$ being the masses of threshold states and $B_{\infty}$ the infinite volume binding energy. Since here $m_{1}, m_{2}$ are masses of two baryons with multiple heavy quarks, $\Delta_{FV}$ receives a large suppression even when $B_{\infty}$ is of a few MeV. This assures that the binding energies for these dibaryons, particularly for $\mathcal{D}_{bc}$, will be very close to the extracted mass differences that we showed in Figure~\ref{fig:DB_qQ}. Since we use the strange, charm and bottom quark masses already at their physical values, the extracted mass differences, after assuming negligible finite volume corrections, are indeed the physical binding energies of these dibaryons. However, to confirm the bound state nature of these dibaryons, particularly for $\mathcal{D}_{bs}$ and $\mathcal{D}_{cs}$, one needs to carry out a detail finite volume study \cite{Luscher:1990ck} and we will address that in future. 

For dibaryons, $\mathcal{D}_{cu}$ and $\mathcal{D}_{bu}$ we first perform chiral extrapolations with a constant plus a term linear in $m_{\pi}^2$. Due to the presence of light quarks, the signal to noise ratios in the correlation functions of these states are rather poor. Moreover for $\mathcal{D}_{bu}$, one decay product is the resonance state $\Delta_{uuu}$, which needs to be addressed with adequate finite volume study \cite{Luscher:1990ck}.  With the current lattice set up, it is therefore difficult for us to make a precise conclusive statement about the stability of these two dibaryons. In future such a study can be carried out with the availability of more computing resources. However, following the example of deuteron, if we assume they are also bound, our results suggest that their binding energies will be much smaller compared to other three dibaryons mentioned above.
%
\bet[t]
\centering
\begin{tabular}{c | c | c | c}\hline \hline 
  Dibaryon & Energy difference & Energy difference & Mass\\
           & from spin-1/2   & from spin-3/2   &   \\
           & threshold [MeV]      & threshold [MeV]       & [MeV]\\ \hline \hline
	$\mathcal{D}_{bc}$ & $-91(12)$   & $-52(13)$  & $19105(21)$ \\ \hline
	$\mathcal{D}_{bs}$ & $-287(45)$ & $-29(13)$  & $16004(17)$ \\ \hline
	$\mathcal{D}_{cs}$ & $ -26(9)$  & $-90(20)$ & $6381(20)$\\ \hline
	$\mathcal{D}_{bu}$ & $-350(110)$ & $3(50)$  & \\ \hline
  $\mathcal{D}_{cu}$ & $-8(17)$ & $-75(46)$ & \\ \hline \hline
  $\mathcal{D}_{bq (m_q = 1.38m_c})$ &  & $-60(10)$& \\ \hline
  $\mathcal{D}_{bq (m_q = 1.72m_c)}$ &  & $-87(8)$ & \\ \hline
  $\mathcal{D}_{bq (m_q =  2.07m_c)}$ &   & $-101(8)$ & \\ \hline
  $\mathcal{D}_{bq (m_q = m_b)}$ &  & $-109(5)$ & \\ \hline \hline
  $\mathcal{D}_{bq (m_q = 2m_b)}$ &  & $-85(10)$ & \\ \hline
  $\mathcal{D}_{bq (m_q = 5m_b)}$ &  & $-15(8)$ & \\ \hline \hline
\end{tabular}
\caption{\label{tb:mass}Energy differences between spin-1 heavy dibaryons and their two-baryon thresholds. The third column shows the predictions of masses for the stable dibaryons. The bottom part of the table is for dibaryons with unphysical quark, $q$, with its mass ($m_q$) varying in between the charm quark mass and five times the bare bottom quark mass. The errors within brackets incorporate both statistical and systematic errors added in quadrature.}
\eet{tb:op}
\bef[b]
\vspace*{-0.1in}
\hspace*{-0.1in}
\includegraphics[scale=0.43]{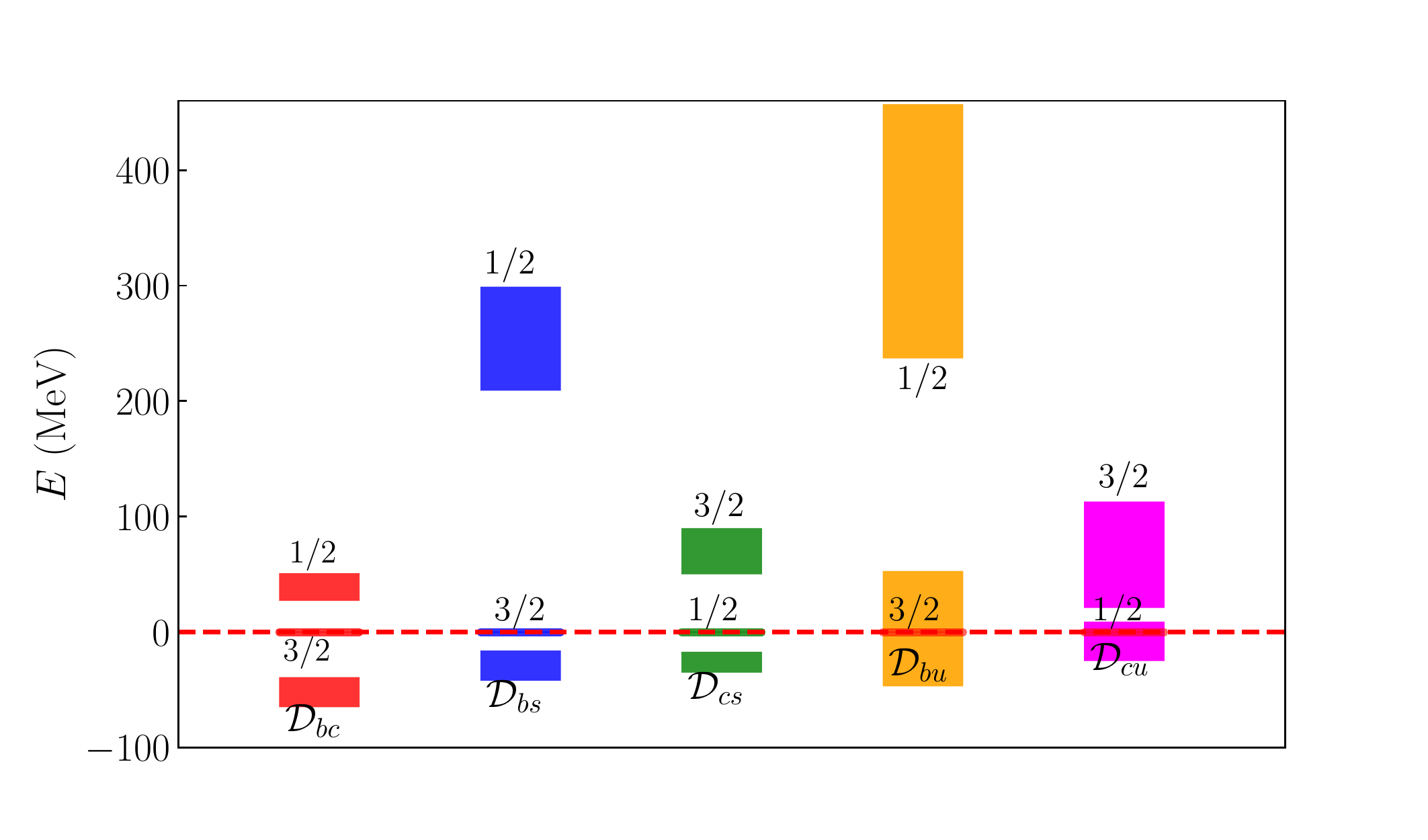}
\vspace*{-0.09in}
\caption{\label{fig:delta_DB}Relative energy levels of the spin-1 dibaryons and their respective two-baryon non-interacting strong-decay threshold states. The horizontal line is for the closest threshold (normalized to zero) while $1/2$ and $3/2$ signify the spin of the constituent single baryons of the two-baryon threshold.}
\eef{}
The final values of dibaryon masses are calculated by adding the known two-baryon threshold masses to the continuum value of binding energies that we extracted. For masses of yet-to-be discovered baryons we use their lattice values as calculated in this work and in Refs. \cite{Brown:2014ena,Mathur:2018epb, Mathur:2018rwu}. We also use subtraction method \cite{Dowdall:2012ab,Brown:2014ena,Mathur:2018epb, Mathur:2018rwu} utilizing spin-average value of $1S$ charmonia and bottomonia and find results consistent with above mass estimates. The final values of dibaryon masses are shown in the third column of Table \ref{tb:mass}. In Fig. \ref{fig:delta_DB} we show the relative energy levels of these dibaryons with respect to their spin-1/2 and spin-3/2 two-baryon non-interacting strong-decay thresholds.
It is quite apparent that the dibaryons $\mathcal{D}_{bc}, \mathcal{D}_{bs}$ and $\mathcal{D}_{cs}$ lie below their respective closest thresholds (horizontal zero line) by about 4, 2 and 3 sigma errors, respectively, while for the other two, $\mathcal{D}_{bu}$, and $\mathcal{D}_{bu}$, we could not reach to a definitive conclusion due to large errors. It may be noted here that we have obtained our results from simulations with a finite number of statistical ensembles, and on space-time grids with finite lattice spacings and finite volume, and hence these are associated with both statistical and systematic errors. Our final results are obtained after carefully addressing those errors and we elaborate that in SI.

Interestingly, our results point out that the strong binding energies of these heavy dibaryons increase as they become heavier in mass. 
To confirm this pattern we also calculate similar dibaryons, $\mathcal{D}_{bq}$, with bottom quark at its physical value while varying the other quark mass ($m_q$) between the charm quark mass to five times the bare bottom quark mass (for $m_q > m_b$, we use the same tuning coefficients as for $m_q = m_b$). Of course, such quarks are unphysical and so are these dibaryons, but since they obey the same strong dynamics they can clarify the pattern of stability. In Table \ref{tb:mass}, at the bottom part, we tabulate the binding energies of these unphysical dibaryons (these results are obtained only at our finest lattice).

To depict it more clearly, in Fig. \ref{fig:DB_e}  we plot these binding energies for dibaryons $\mathcal{D}_{bq}(\Omega_{bbb}\Omega_{qqq})$ with $q$ varying from light to all the way to the bottom quark. It clearly shows that the spin-1 dibaryons become more stable when they are heavier. We even calculate such dibaryons with much higher quark masses (which are of course unphysical) and observe that their binding energies decrease when $m_q > m_b$ and they vanish at very large quark masses ($m_q \rightarrow \infty$). This suggests that the combination $\mathcal{D}_{bq}(\Omega_{bbb}\Omega_{qqq,q=b})$ has the maximum binding.

\bef[t]
\centering
\includegraphics[scale=0.44]{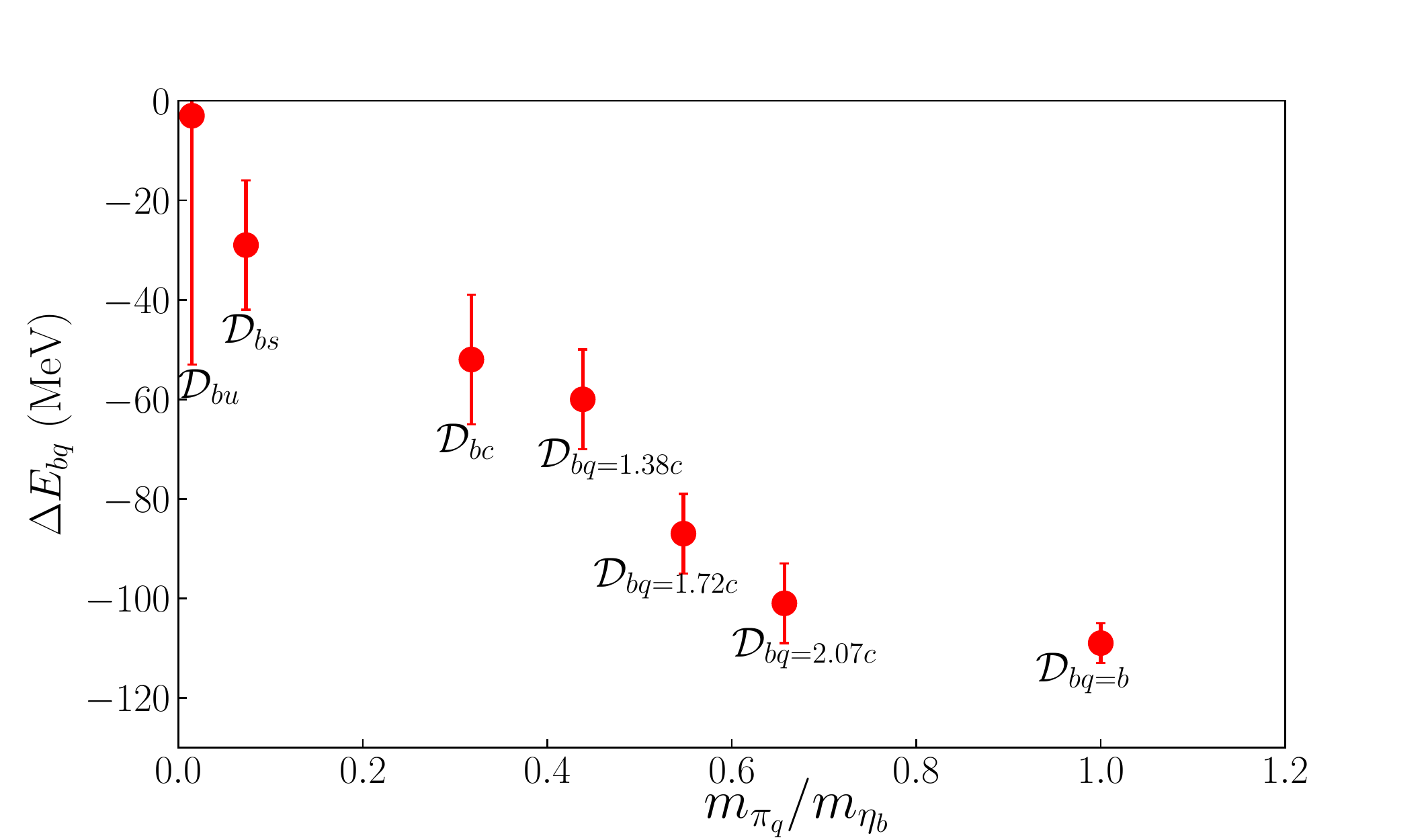}
\vspace*{-0.09in}
\caption{\label{fig:DB_e} Binding energy of the spin-1 dibaryon $\mathcal{D}_{bq}$ where we kept the bottom quark fixed and vary the other quark, $q$, from light to bottom quarks. Along the $x$-axis we show the ratio of the pseudoscalar mass at $m_q$ ($m_{\pi_q}$) to that at the bottom quark ($m_{\eta_b}$). As the mass of the quark $q$ increases ($m_q = m_u$ to $m_b$), the dibaryon binding energy also increases.}
\eef{}

The strong-binding of these dibaryons and the pattern that it
increases with the increase of quark mass is pointing towards an
interesting strong dynamics in the heavy hadrons. It is to be noted
that for the doubly heavy tetraquarks lattice studies \cite{Francis:2016hui,Junnarkar:2018twb,Francis:2018jyb,Leskovec:2019ioa} found a
complete opposite patterns where the strong-binding increases as the
light quark becomes lighter. This suggests that the
strong-binding mechanisms of spin-1 heavy tetraquarks and spin-1 heavy
dibaryons are different and certainly this needs to be investigated
further, perhaps similar to the way as investigated in Ref. \cite{Francis:2018jyb} for the heavy tetraquarks.

The stability of these dibaryons against the strong and electromagnetic decays opens up the possibility of finding these and even more complicated higher nuclei with many heavy quarks, similar to nuclei of various elements in our periodic table. Similar to the role of deuteron in the nuclear fusion cycle for the creation of elements, these dibaryons can help the fusion processes of heavy baryons to produce nuclei with heavy quarks. Such nuclei can be studied theoretically in future with adequate computational resources and it may well be possible to discover them in future higher energy heavy ion facilities. It will also be interesting to see if the formation of such states has any implication in cosmology.
Formation of such hadrons also enhances the speculation on the possibility of a quark-level analogue of nuclear fusion which was discussed recently in Ref. \cite{Karliner:2017elp}. For example, formation of $\mathcal{D}_{bs}$ through fusion of $\Omega_{bb}$ and $\Omega_{b}$, as well as through fusion of $\Omega_{bbb}$ and $\Omega_{sss}$, are highly exothermic with the release of energy about 300 and 30 MeV, respectively. We also find that reactions such as $\mathcal{B}^{1/2}_{qbb} + \mathcal{B}^{1/2}_{qqb} \rightarrow \mathcal{B}^{3/2}_{bbb} + \mathcal{B}^{3/2}_{qqq}, q\equiv c,s,u/d$, and
$\mathcal{B}^{3/2}_{ccc} + \mathcal{B}^{3/2}_{qqq} \rightarrow \mathcal{B}^{1/2}_{qcc} + \mathcal{B}^{1/2}_{qqc}, q\equiv s,u/d$, are highly exothermic (here we represent $\mathcal{B}^Jq_1q_2q_3$ as a baryon with spin $J$ and quark contents $q_1,q_2$ and $q_3$).

Since these dibaryons involve quarks with masses over a wide range, studies of their properties will be helpful to understand the strong dynamics at multiple scales.
Due to the presence of multiple heavy quarks they will decay via various possible weak decay processes. For example, $\mathcal{D}_{cb}$ can decay through $b \rightarrow c$, $b \rightarrow s$ and $c \rightarrow s$ to various light baryons and multiple  mesons which can interfere among themselves. Detailed analysis of these multiple ways of decay may also be an ideal place to study the hadronic-interference in weak decay processes.
Furthermore, this work, particularly the strong-binding of heavier dibaryons, motivate us to study other heavy dibaryons, namely spin-0 states as well as heavy quark analogues of the H-dibaryon. Results on those dibaryons will be reported in future which, taking together with the finding here,  may well be able to infer important information on the dynamics and binding mechanisms of heavy hadrons and perhaps also for the nuclei with heavy quarks.

{\bf{Acknowledgements:}}
 Computations were carried out on the Cray-XC30 of ILGTI, TIFR, 
and on the Gaggle/Pride clusters of the Department of Theoretical Physics,
TIFR. We would like to thank A. Dighe, M. Padmanath, S. Pavaskar and S. Raychaudhuri for many useful discussions.
We are thankful to the MILC collaboration 
and in particular to S. Gottlieb for providing us with the HISQ lattices.   N. M. would also like to thank A. Salve, K. Ghadiali and P. M. Kulkarni for computational support. 

\bibliography{hdb}

\section*{Supplementary Information}
Below we detail the procedure that we utilize for computing energy differences between the ground state dibaryons from their non-interacting two-baryon thresholds. Afterwards we address statistical and systematic errors associated with this calculation.
\\

\noindent{\bf{Calculating mass differences and masses:}} \,\,  
Euclidean two-point correlators for single and dibaryons are calculated using wall source smearing and from those we extract their ground state masses. We calculate the mass difference that we showed in Table \ref{tb:mass} as,
\begin{equation}
\label{Eq:eff_split}
\Delta E  = M_{\mathcal{D}} - M_{B_1B_2},
\end{equation}
where $M_{\mathcal{D}}$ is the ground state mass obtained from the two-point dibaryon correlators, while $M_{B_1B_2} = M_{B_1} + M_{B_2} $ is the mass of the non-interacting two-baryon ($B_1$ and $B_2$) threshold. We calculate this mass difference by two methods: first by fitting the single baryons and dibaryon data sets separately and then computing their difference in Jackknife sample. In the second method we take the Jackknife ratio of the dibaryon correlator, $\mathcal{D}(\tau)$, to the two-baryon correlators, $B_1(\tau) \times B_2(\tau)$) as: 
\begin{equation}\label{Eq:approach_4}
\mathcal{D}^\prime (\tau) = \frac{\mathcal{D}(\tau)}{B_1(\tau) \times B_2(\tau)} \rightarrow \mathcal{A} e^{-\Delta E \tau} + ...
\end{equation}
A fit to the ratio correlator ($\mathcal{D}^\prime(\tau)$) then directly yields  the mass splitting with respect to the relevant threshold ($M_{B_1B_2}$). However, one must be careful in using ratio method as it can produce spurious effects due to saturation of different particle at different time slices. Results from both methods are found to be consistent in our case.
\\

\noindent{\bf{Continuum extrapolation:}} \,\,
Due to exact chiral symmetry on lattice, overlap fermions have no $\mathcal{O}^n(ma), n = 1, 3, 5, \cdots$ errors. Hence the first two terms that enter in chiral extrapolations are $a^2$ and $a^2\rm{ln}(a)$. Since we have only three lattice spacings it is impossible to fit them together and so we fit them separately. However, with the given precision and with data at the finest lattice spacing it is not possible to distinguish two fittings with those two forms. We also fit them in a constrained fitting similar to those in Ref. \cite{Lepage:2001ym}. 
\\

\noindent{\bf{Final results on dibaryon masses:}} \,\,
The final values of dibaryon masses are calculated using two different subtraction methods. In the first method we add the continuum extrapolated mass differences with the two-baryon threshold masses. For example, for $\mathcal{D}_{bc}$ we add the mass difference of $52(13)$ MeV with the spin-3/2 $\Omega_{bbb}$ and $\Omega_{ccc}$ masses. For single baryons, which have not been discovered yet, we have calculated their masses and found those to be consistent with Refs. \cite{Mathur:2018epb, Mathur:2018rwu,Brown:2014ena}. In the second method, following Refs. \cite{Mathur:2018epb, Mathur:2018rwu}, we first calculate subtracted masses on the lattice as
\beq
\Delta M_{\mathcal{D}} = [M^{L}_{\mathcal{D}} - n_c\overline{1S}_c/2 - n_b\overline{1S}_b/2]a^{-1}.
\eeq{submass}
Here $n_c$ and $n_b$ are the number of $c$ and $b$ valence quarks in dibaryons
and  $\overline{1S}_c$ and $\overline{1S}_b$ are the lattice calculated spin average $\overline{1S}$ charmonia and bottomonia masses respectively. These subtracted masses are then extrapolated to their continuum limit ($\Delta M^c_{\mathcal{D}}$) and finally the physical results are obtained by adding physical values of spin average masses to that as
\begin{equation}
  M^c_{\mathcal{D}} = \Delta M^{c}_{\mathcal{D}} + n_b(\overline{1S}_c)_{phys}/2 + n_c(\overline{1S}_b)_{phys}/2.
\end{equation}
We find results from both methods are consistent with each other.\\

\noindent{\bf{Error analysis:}} \,\,
Results obtained at finite volume lattices have both statistical as well as systematic errors. We address each of those below which are similar to the error analysis in Refs. \cite{Mathur:2018epb, Mathur:2018rwu,Brown:2014ena}.

\noindent{\it{Statistical error}}: We use wall source to obtain dibaryon correlators and it helps to obtain better ground state plateau. We find a statistical uncertainty of 10 MeV while calculating mass difference for $\mathcal{D}_{bc}$.

\noindent{\it{Fitting window error}}:
With stable plateau we find uncertainty due to different {\it fitting windows} to be maximum of 4 MeV.

\noindent{\it Discretization error}: To obtain reliable results for these dibaryons with heavy quarks, a crucial issue is the control of discretization errors. The following three methods help us to reduce discretization errors: $i$) continuum extrapolation, as mentioned above, of the results obtained at three lattice spacings, the finest one being at 0.06 fm, $ii$) use of overlap action and $iii$)  extraction of mass differences, which has less discretization error than masses. As mentioned previously, the first two terms that enters in the continuum extrapolation are $a^2$ and $a^2\,\rm{ln}(a)$. With three data points we fit these two forms separately and also together in a constrained fitting. Differences in central values at the continuum limit obtained with these different fittings are included in the discretization errors. After continuum extrapolation, we find the maximum discretization would be less than 4 MeV for the mass difference in $\mathcal{D}_{bc}$. 

\noindent{\it Scale setting error}: On these set of ensembles we have also determined scales by measuring the $\Omega_{sss}$
baryon mass and those were found to be consistent with the determinations using $r_1$ parameter \cite{Basak:2013oya}. Measurement of scale with Wilson flow~\cite{Bazavov:2015yea} was also found to be consistent with the scale used here.
The scale setting uncertainty in the mass difference (Eq. (2)) for $\mathcal{D}_{cs}$ is found to be $\sim$ 4 MeV.

\noindent{\it Quark mass tuning error}: We use the Fermilab method of heavy quarks \cite{ElKhadra:1996mp} to tune the charm and bottom quark masses.  In this method, we calculate the kinetic masses, instead of pole masses, of the spin average $\overline{1S}$ quarkonia and equate those with their experimental values. We perform this process corresponding to the central value of the scale and also with central $\pm$ error values. For each of the tuned masses, we calculate hadron masses and include the variation as the errors due to quark mass tuning. Our estimate for the mass tuning errors in the energy difference for $\mathcal{D}_{bc}$ due to the charm and bottom quarks are about 2-3 MeV.

\noindent{\it Finite size effects}: The finite volume corrections, $\Delta_{FV}$, for these multi-hadron systems were found to be $\sim {\mathcal{O}}(e^{-k_{\infty} L})/L$, where $k_{\infty} = \sqrt{(m_1+m_2)B_{\infty}}$, $m_1, m_2$ being the masses of threshold states and $B_{\infty}$ infinite volume binding energy. Since here $m_{1}, m_{2}$ are masses of two heavy baryons, $\Delta_{FV}$ would be very small even if $B_{\infty}$ is of a few MeV. Following our works on heavy baryons in Ref. \cite{Mathur:2018epb} and similar works in Ref. \cite{Brown:2014ena}, we include an uncertainty of 2 MeV from finite volume effects while calculating the mass difference for $\mathcal{D}_{bc}$. However, the fine volume effects for dibaryons with light-quarks would be bigger and one needs to perform a dedicated finite volume study for that \cite{Luscher:1990ck}.\\

\noindent{\it Other sources}: For dibaryons $\mathcal{D}_{cu}$ and $\mathcal{D}_{bu}$, we had to carry out chiral extrapolations and we perform that with constant plus $m_{\pi}^2$ terms. However, since the pion masses are not so light, particularly on the fine lattice, it is difficult to control the systematic associated with the chiral extrapolation. The dibaryon $\mathcal{D}_{bu}$ also involves its decay to $\Delta$-resonance which needs to be treated with finite volume study and that is beyond the scope of this work.
We have thus included large errors associated with it coming from chiral and possible finite volume corrections which in turn limits the precision of our predictions for these baryons. In future with the availability of more computer resources this limitation could be addressed. 
No chiral extrapolation is necessary for $\mathcal{D}_{cs}$, $\mathcal{D}_{bs}$ and $\mathcal{D}_{bc}$ for which we have made conclusive prediction.

The unphysical sea quark mass effects are expected to be within 
a percent level for heavier dibaryons with no effective valence light quark content~\cite{McNeile:2012qf, Dowdall:2012ab, Chakraborty:2014aca}. This effect however would be larger for $\mathcal{D}_{cu}$ and $\mathcal{D}_{bu}$ dibaryons.
Errors due to mixed action effects are found to be small within this lattice set up~\cite{Basak:2014kma}
and are expected to vanish in the continuum limit. 
Errors from electromagnetism are expected  to be within 3 MeV for baryons~\cite{Borsanyi:2014jba} and should be similar for these dibaryons.

Adding all these systematic errors in quadrature, we found that for the energy difference (Eq. 2), i.e., for the binding energy, in the case of $\mathcal{D}_{bc}$ is less than 8 MeV. For other dibaryons we also estimated those similarly and included those in table II. We find that systematic errors for dibaryons  $\mathcal{D}_{cu}$ and $\mathcal{D}_{bu}$  are too big to reach a definitive conclusion about their bindings.


Below we summarize the error budget for $\mathcal{D}_{bc}$ dibaryon.
\bet[h]
\vspace*{-0.07in}
\centering
\begin{tabular}{l|c }
$Source$ & Error (MeV)\\
\hline
Statistical & 10  \\
 Discretization & 5  \\
  Scale setting& 4   \\
  $m_b$ tuning & 3  \\
  $m_c$ tuning & 3 \\
  Fit window & 4 \\
  Finite size & 2 \\
  Electromagnetism & 3\\
  \hline
  Total & 10 (stat) \& 9 (syst)\\
  \hline
\end{tabular}
\caption{Error budget in the calculation of energy splittings for $\mathcal{D}_{bc}$ dibaryon.}
\end{table}

\end{document}